\documentclass[conference]{IEEEtran}
\IEEEoverridecommandlockouts

\usepackage{cite}
\usepackage{amsmath,amssymb,amsfonts}
\usepackage{algorithm}
\usepackage{algorithmic}
\usepackage{graphicx}
\usepackage{textcomp}
\usepackage{xcolor}
\usepackage{xurl}
\usepackage{hyperref}
\def\BibTeX{{\rm B\kern-.05em{\sc i\kern-.025em b}\kern-.08em
    T\kern-.1667em\lower.7ex\hbox{E}\kern-.125emX}}
\begin{document}

\title{Runtime-Orchestrated Second-Order Optimization for Scalable LLM Training}


\author{
\IEEEauthorblockN{Yishun Lu, Junhao Zhang, Zeyu Yang, and Wes Armour}
\IEEEauthorblockA{\textit{Engineering Science Department} \\
\textit{University of Oxford}\\
Oxford, United Kingdom \\
\{yishun.lu, junhao.zhang, zeyu.yang\}@eng.ox.ac.uk, wes.armour@oerc.ox.ac.uk}
}
\maketitle

\begin{abstract}
Second-order methods offer an attractive path toward more sample-efficient LLM training, but their practical use is often blocked by the systems cost of maintaining and updating large matrix-based optimizer states. We introduce \textbf{Asteria}, a runtime system designed to remove this bottleneck by separating second-order optimization logic from the critical GPU training path. Rather than keeping all preconditioner state on the accelerator, Asteria dynamically distributes optimizer state across GPU memory, CPU memory, and optional NVMe storage according to architectural constraints and runtime pressure. It further uses training hooks to prepare shadow states in advance, allowing expensive inverse-root computations to proceed asynchronously on the host while GPU computation continues. For distributed training, Asteria employs a bounded-staleness protocol that limits synchronization frequency while preserving optimizer effectiveness through topology-aware coordination. 

We evaluate Asteria on both memory-constrained and distributed training settings. On a DGX Spark platform with a single GB10 GPU and 128GB unified memory, Asteria supports second-order training for a 1B-parameter language model. On multi-node GH200 systems, it lowers visible optimizer overhead, reduces recurring latency spikes, accelerates convergence in wall-clock time, and maintains the optimization advantages of SOAP and KL-Shampoo. Our results suggest that second-order LLM training can be made practical not by simplifying the optimizer alone, but by rethinking how optimizer state, background computation, and distributed synchronization are managed at the runtime level.
\end{abstract}

\section{Introduction}

The exponential scaling of Large Language Models (LLMs) and their associated large amount of pre-training corpora has pushed modern deep learning clusters to their operational limits. Currently, large-scale training is dominated by first-order optimizers, such as AdamW \cite{loshchilov2019decoupledweightdecayregularization}. While highly scalable, these methods suffer from inherent statistical inefficiency, requiring massive token volumes to reach convergence. Recent advances in second-order optimization \cite{shampoo_raw, vyas2025soap,eschenhagen2026clarifyingshampooadaptingspectral,lin2026understandingimprovingshampoosoap} suggest new approaches to address this statistical bottleneck, demonstrating significantly accelerated convergence rates and lower asymptotic training loss.

Despite these algorithmic advantages, deploying second-order methods at LLM scale introduces severe system-level barriers. These methods must continuously accumulate empirical gradient covariance statistics to construct structural preconditioners that capture the curvature of the loss landscape. Doing so requires storing large covariance matrices ($L$ and $R$) and repeatedly computing dense eigendecompositions. The result is a prohibitive $\mathcal{O}(N^3)$ computational cost and an $\mathcal{O}(N^2)$ memory footprint. At extreme scale, second-order optimization therefore demands vast, highly aggregated memory resources. The challenge becomes even more severe on memory-constrained edge nodes and workstations. In Unified Memory Architectures (UMA), the GPU and CPU share a fixed physical memory budget. Large optimizer states can rapidly exhaust this pool, directly competing with activations for memory capacity. This creates a harsh trade-off: users must either shrink model capacity and batch size, or risk catastrophic Out-of-Memory (OOM) failures as model size grows. As a result, the second-order paradigm becomes less practical for local deployment for large models.

State-of-the-art implementations, such as \texttt{Distributed Shampoo} ~\cite{shi2023distributeddataparallelpytorchimplementation} and ASDL \cite{osawa2023asdlunifiedinterfacegradient}, attempt to mitigate this by utilizing \texttt{DTensor} to horizontally shard the optimizer states across multiple GPUs, synchronizing via a global \texttt{AllGather} at the end of each iteration. However, this design remains trapped within the paradigm of VRAM and homogeneous flat communication, inevitably colliding with three physical walls in real-world heterogeneous and extreme-scale environments:

\begin{itemize}

    \item \textbf{The Vertical Capacity Wall:} Existing frameworks rely heavily on multi-GPU sharding to amortize optimizer-state memory costs. Once deployment degrades to a resource-constrained single-node environment (e.g., a Nvidia DGX Spark workstation with a single GB10 GPU and 128GB of unified memory, or a Nvidia RTX3090 with 24GB VRAM), horizontal sharding is no longer available. The resulting monolithic second-order states must reside within a fixed local memory budget, and can therefore exceed the physical capacity of the device, triggering catastrophic Out-of-Memory (OOM) failures.

    \item \textbf{The Overlap Disruption Wall:} In distributed training systems such as FSDP \cite{zhao2023pytorchfsdpexperiencesscaling}, high efficiency depends on maintaining fine-grained overlap between computation and parameter communication. Existing second-order methods often violate this requirement. While they attempt to amortize costs through algorithmic relaxation (e.g., updating preconditioners at fixed frequencies like $pf=10$), this merely delays rather than eliminates the bottleneck. These monolithic updates commonly require cubic-cost matrix inversions or eigendecompositions with complexity $\mathcal{O}(d^3)$, where $d$ denotes the dimension of the corresponding preconditioner matrix. Such updates break the intended compute--communication overlap, stall the execution pipeline, and leave the GPU idle with severe periodic latency spikes.

    \item \textbf{The Global Consensus Wall:} While modern communication primitives (e.g., NCCL) can efficiently map collectives onto hierarchical hardware topologies, existing second-order optimization algorithms still rigidly require synchronous global state updates across all participating ranks. This strict mathematical consensus tightly couples fast intra-node communication (e.g., NVLink/PCIe) with slow inter-node synchronization (e.g., InfiniBand), thereby preventing the system from exploiting asynchronous, topology-aware staleness bounds.

\end{itemize}

To overcome the deadlock in which preconditioner states are too large for a single GPU but too expensive to communicate across multiple GPUs, we introduce \textbf{Asteria}, a hardware-software co-designed runtime for extreme-scale second-order optimization. Asteria is highly elastic: it can scale down to operate within strict single-GPU memory budgets, and scale up to support large FSDP clusters efficiently. It addresses the three physical walls above through the following core novelties:

\begin{itemize}
    \item \textbf{Architecture-Adaptive Asymmetric Memory Tiering:}
    Asteria introduces an architecture-adaptive memory placement strategy for second-order optimizer state that is tailored to the asymmetric lifecycle of matrix preconditioning. Instead of treating optimizer state as homogeneous tensors to be uniformly offloaded, Asteria separates where second-order states reside according to how they are produced and consumed: Kronecker factors are updated on the GPU, inverse-factor states are maintained in CPU-accessible memory, and the expensive inverse-root computation is executed asynchronously on CPU-side snapshots before the resulting states are exposed back to the GPU for consumption. On unified-memory platforms, this design leverages CPU-preferred managed memory with direct GPU mapping to avoid redundant host-to-device replication of updated inverse factors. On capacity-limited runs, it further extends the same lifecycle-aware policy with asynchronous NVMe-backed staging and explicit reclamation. This novelty is therefore not NVMe offloading alone, but a second-order-specific memory hierarchy that matches the asymmetric compute-and-access pattern of matrix preconditioning. 

    \item \textbf{Hook-Orchestrated Shadow-State Pipeline:}
    Asteria removes the expensive $\mathcal{O}(d^3)$ inverse-root update from the latency-critical GPU training path by dispatching it asynchronously to CPU worker threads, while the main FSDP forward and backward passes continue with bounded-staleness preconditioner states. To stage these updated states back toward the GPU without intrusive changes to the training stack, Asteria repurposes lightweight FSDP forward and backward hooks as scheduling signals for a low-priority shadow pipeline that drains ready state transfers and issues prefetches for soon-to-be-consumed inverse factors. This design uses hooks, bounded staging, and asynchronous state preparation to exploit slack in the execution timeline and reduce synchronization-induced stalls on the critical path. This marks a paradigm shift from algorithmic amortization to true system-level decoupling, completely flattening the periodic $\mathcal{O}(N^3)$ latency spikes to achieve a smooth hardware execution profile parallel to first-order methods.
    
    \item \textbf{Bounded-Staleness Selective Coherence for Host-Resident Second-Order State:}
    Asteria replaces rigid full-state synchronization with a bounded-staleness coherence protocol for host-resident second-order preconditioners. Rather than synchronizing every block at every step, Asteria tracks per-block freshness and communicates only those blocks whose staleness budget has been exceeded. These selected blocks are synchronized directly in CPU-accessible memory through node-aware hierarchical process groups, first within each node and then across node representatives before being propagated back to local peers. While such host-side communication backends may offer lower peak bandwidth than GPU-resident collectives, Asteria sharply reduces the total communication volume by sparsifying synchronization at block granularity and by avoiding unnecessary host-device round trips. 

\end{itemize}

We evaluate Asteria across both scale-up and scale-out regimes. On a memory-constrained Nvidia DGX Spark system (single GB10, 128GB unified memory), Asteria is able to initialize and train a 1B-parameter LLM with second-order optimization, demonstrating that second-order state can be kept within a fixed single-node memory envelope without relying on horizontal multi-GPU sharding. On a multi-node Nvidia GH200 cluster, step-time profiling shows that Asteria suppresses the periodic latency spikes of native second-order execution and keeps the training trajectory close to that of first-order methods in wall-clock time while preserving the favorable loss descent of second-order optimization. Hardware telemetry further shows that this systems-level decoupling improves the energy-loss tradeoff: Asteria reduces the exposed cost of second-order state maintenance, lowers total energy relative to native second-order baselines, and improves the amount of loss reduction achieved per unit of energy. Together, these results show that Asteria makes second-order optimization practical under tight memory budgets and more effective in wall-clock time and energy at distributed scale.

\section{Related Work}

\subsection{Distributed Second-Order Optimization}

Recent years have seen renewed interest in second-order optimization because of its statistical efficiency. Methods such as K-FAC \cite{martens2020optimizingneuralnetworkskroneckerfactored}, SOAP \cite{vyas2025soap}, and Shampoo \cite{shampoo_raw}, together with their variants \cite{anil2021scalablesecondorderoptimization, shi2023distributeddataparallelpytorchimplementation, eschenhagen2026clarifyingshampooadaptingspectral, lin2026understandingimprovingshampoosoap, lu2026meanfisherorthogonalprojectionnatural}, show that curvature-aware updates can accelerate convergence. AdamW \cite{loshchilov2019decoupledweightdecayregularization} uses diagonal preconditioning, where each parameter is scaled independently. In contrast, second-order methods use matrix-structured curvature information to precondition updates. In Shampoo \cite{shampoo_raw}, a matrix-shaped gradient $G_t$ is preconditioned by Kronecker-factor inverse roots:
\begin{equation}
\tilde G_t = L_t^{-1/4} \, G_t \, R_t^{-1/4}.
\end{equation}

In SOAP \cite{vyas2025soap}, the gradient is first projected into an orthogonal basis induced by second-order statistics, adaptively scaled in that basis, and then mapped back:
\begin{equation}
\tilde G_t = Q_L \, \phi\!\left(Q_L^\top G_t Q_R\right) Q_R^\top,
\end{equation}
where $\phi(\cdot)$ denotes coordinate-wise adaptive scaling in the transformed basis.

A common design principle in these methods is the use of structured approximations, including Kronecker factorization, to reduce the cost of second-order preconditioning. 
Among large-scale implementations, PyTorch Distributed Shampoo \cite{shi2023distributeddataparallelpytorchimplementation} is a representative systems baseline. Its main strategy is to shard large preconditioner states across multiple GPUs using \texttt{DTensor}, which reduces the memory footprint on each device. This design is effective when aggregate GPU memory is sufficient, and it reflects a common assumption in prior systems for second-order optimization: optimizer state remains primarily in GPU memory and is managed through distributed in-device sharding.

This assumption becomes limiting in scale-up settings with restricted device parallelism or tight memory budgets. On systems with a single GPU, horizontal sharding is not available, and the full second-order state must fit within a single device-visible memory domain. Under this setting, existing second-order implementations may fail at initialization or early in training because of memory exhaustion. In addition, current software stacks are largely built around high-level Python execution, which limits access to low-level mechanisms for state placement, asynchronous storage staging, and explicit memory reclamation.

\subsection{Heterogeneous Memory Offloading for ML}
Heterogeneous memory offloading has become a standard approach for training large models under limited GPU memory. Systems such as ZeRO-Offload \cite{ren2021zerooffloaddemocratizingbillionscalemodel}, ZeRO-Infinity \cite{rajbhandari2021zeroinfinitybreakinggpumemory}, and PatrickStar \cite{Fang_2023} extend the effective memory capacity of training by moving optimizer states, gradients, and parameters across GPU memory, CPU memory, and NVMe storage.

These systems are developed primarily for first-order optimizers such as Adam \cite{kingma2017adammethodstochasticoptimization}, Adafactor ~\cite{shazeer2018adafactoradaptivelearningrates} and AdamW \cite{loshchilov2019decoupledweightdecayregularization}, whose states are dominated by flat vector-valued tensors and element-wise updates. Second-order optimizers introduce a different workload. In Shampoo, the dominant states are structured matrices, and state maintenance includes matrix-valued inverse-root or eigendecomposition computations whose access patterns and compute requirements differ from those of first-order methods. As a result, offloading strategies designed for homogeneous vector states do not transfer directly to second-order preconditioners.

\subsection{Computation-Communication Overlap in Distributed Training}
Modern distributed training frameworks, including PyTorch FSDP \cite{zhao2023pytorchfsdpexperiencesscaling}, Megatron-LM \cite{shoeybi2020megatronlmtrainingmultibillionparameter}, and BytePS \cite{10.5555/3488766.3488792}, improve hardware utilization by overlapping computation and communication during training. In these systems, forward and backward matrix operations are scheduled together with collective communication primitives such as \texttt{AllGather} and \texttt{ReduceScatter}. This execution model is standard in large-scale training.

Under such pipelines, additional optimizer-state movement must be integrated with care because communication resources are already shared by parameter and gradient traffic. This issue is more pronounced for second-order methods, whose preconditioner states introduce extra staging and synchronization costs beyond those of first-order optimizers \cite{osawa2023asdlunifiedinterfacegradient,shi2023distributeddataparallelpytorchimplementation}. Existing second-order implementations commonly perform preconditioner maintenance and state synchronization in a relatively coarse-grained manner, which can increase step-time overhead.

Asteria is designed for this setting. It does not modify the main execution graph of the underlying training framework. Instead, it uses lightweight runtime hooks to trigger auxiliary state staging and prefetch alongside the existing training schedule. Furthermore, inspired by recent advancements in heterogeneous memory management for first-order optimizers \cite{Maurya_2024}, expensive inverse-root updates are executed asynchronously on CPU worker threads outside the main GPU path. This design reduces the exposed overhead of second-order state movement while remaining compatible with existing distributed training frameworks.

\begin{figure*}[h]
    \centering
     \includegraphics[width=1\linewidth]{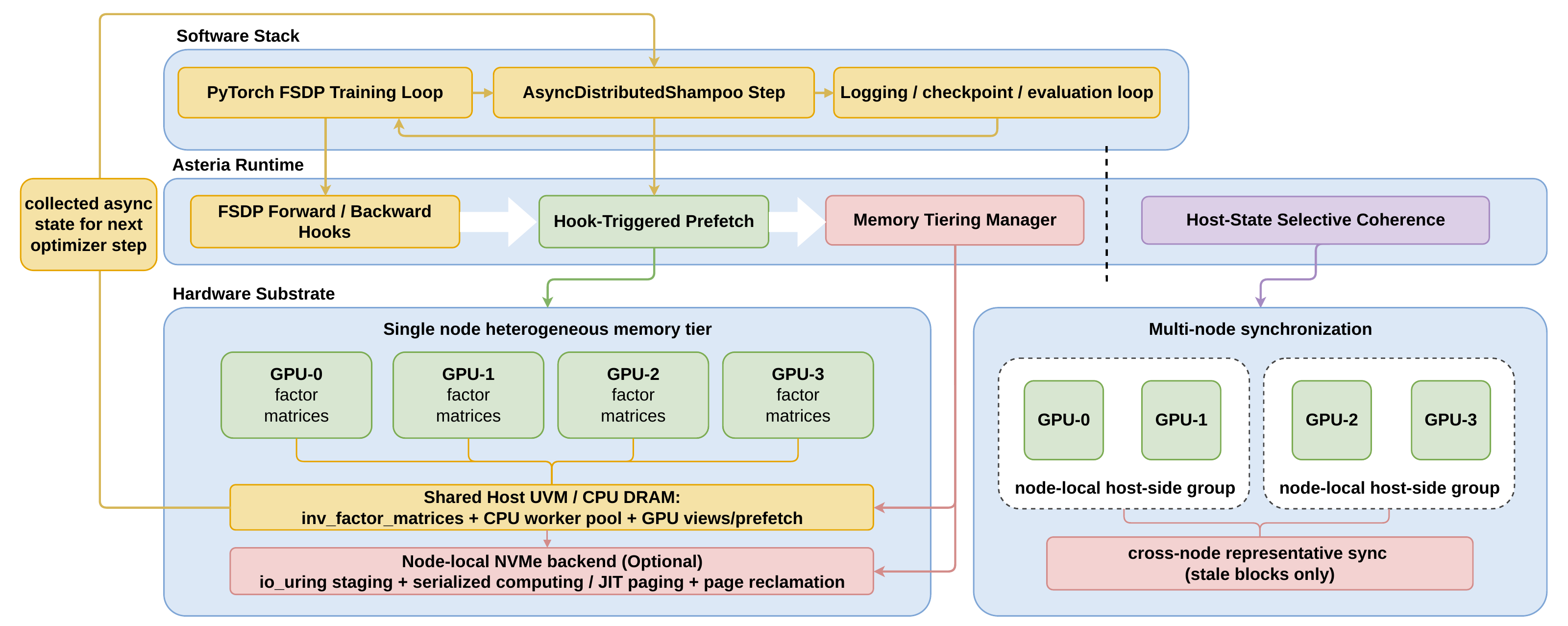}
    \caption{System architecture overview.}
    \label{fig:architecture}
\end{figure*}

\subsection{Relaxed Consistency and Staleness Control}
Many machine learning algorithms can tolerate limited staleness in weights, gradients, or optimizer state. This property is often used to reduce synchronization and communication cost. For example, Local SGD \cite{stich2019localsgdconvergesfast} and GossipGrad \cite{daily2018gossipgradscalabledeeplearning} allow delayed synchronization across workers. In the same spirit, periodic second-order methods \cite{anil2021scalablesecondorderoptimization,vyas2025soap,shi2023distributeddataparallelpytorchimplementation} reduce computation by recomputing inverse preconditioners only once every $N$ steps.

Most of these approaches are topology-agnostic. They reduce synchronization frequency or allow stale state, but they do not adapt the synchronization policy to the structure of the underlying communication hierarchy. In addition, they do not focus on host-resident second-order state, whose placement and communication path differ from those of GPU-resident gradients and parameters.


\section{System Design and Methodology}
\subsection{System Overview}
Asteria is built around state decoupling and runtime scheduling. It operates as a runtime layer around PyTorch FSDP for distributed second-order optimization. Asteria does not modify the main FSDP execution graph. The forward and backward passes remain on the original training path, while second-order state maintenance is handled on a separate runtime path.

As shown in Figure~\ref{fig:architecture}, Asteria separates execution into a foreground training path and an auxiliary path for second-order state. The foreground path covers FSDP forward and backward computation, parameter sharding, and gradient communication. The auxiliary path handles asynchronous inverse-root computation and state staging across GPU memory, host-accessible memory, and an optional NVMe tier. This separation keeps dense $\mathcal{O}(N^3)$ inverse-root updates away from the main GPU path and allows state staging and GPU-visible refresh to proceed alongside training. A bounded-staleness rule determines when the runtime must wait for pending asynchronous inverse-root updates before continuing.

\subsection{Heterogeneous Memory Tiering Tailored for Second-Order State}

Memory offloading is widely used in large-model training systems such as ZeRO-Offload \cite{ren2021zerooffloaddemocratizingbillionscalemodel}. These systems are designed mainly for first-order optimizers such as AdamW, whose states are dominated by flat vector-valued tensors and whose updates are based on element-wise operations. This structure makes first-order state relatively easy to move across GPU memory, CPU memory, and storage tiers.

\begin{figure}[htbp]
    \centering
    \includegraphics[width=1\linewidth]{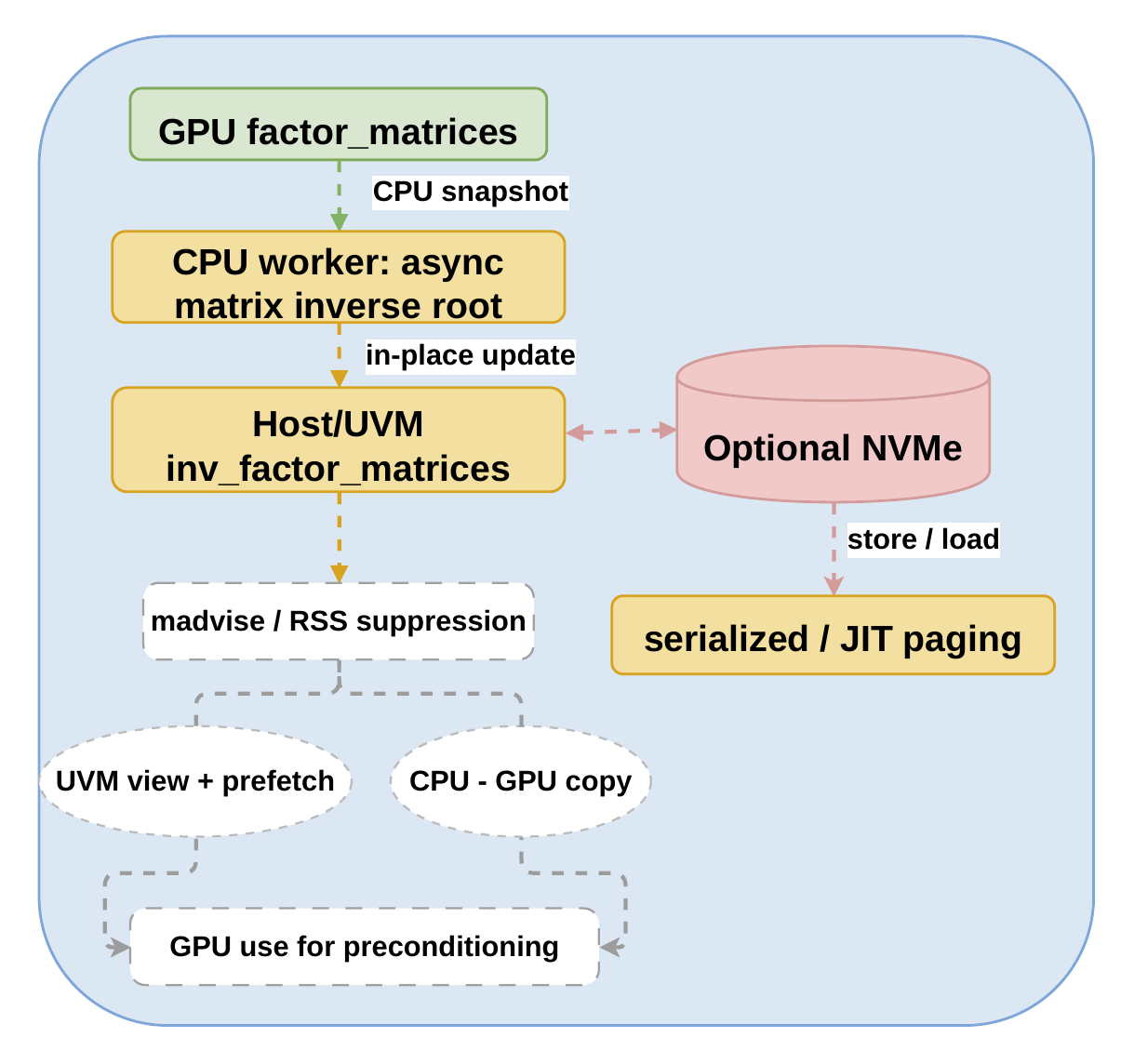}
    \caption{Memory tiering path for second-order state in Asteria.}
    \label{fig:heterogeneous-memory-tiering}
\end{figure}

Second-order methods such as Shampoo \cite{shampoo_raw} impose a different compute and memory pattern. They maintain matrix-structured statistics and require inverse-root updates over large covariance matrices. In Asteria, these inverse-root updates are executed by asynchronous CPU workers, while the training path continues on the GPU. Second-order state management is therefore not only a capacity problem but also a state-placement problem, because different tensors are updated and consumed on different processors.

Figure \ref{fig:heterogeneous-memory-tiering} shows the default memory-tiering path for these states. Asteria keeps \texttt{factor\_matrices} on the GPU, where they are updated during training. Before inverse-root computation, the runtime captures host-side snapshots of these factor matrices and dispatches asynchronous CPU workers to compute updated inverse factors. The updated inverse factors are written back in place to host-accessible \texttt{inv\_factor\_matrices}, which reside in UVM-backed (Unified Virtual Memory) memory or CPU DRAM. When NVMe staging is enabled, these inverse-factor states can be moved through an optional node-local NVMe tier via serialized computing or backward JIT paging. After inverse-root computation or NVMe write-back, Asteria may reclaim unused host pages through \texttt{madvise(MADV\_DONTNEED)} to reduce the resident memory footprint. When the GPU needs the updated preconditioner state, the runtime exposes it through GPU-visible paths that match the memory mode. Under UVM (Unified Virtual Memory), Asteria uses CUDA-accessible views and prefetch. Under non-UVM configurations, it refreshes GPU-side cached copies explicitly.

\begin{figure}[h]
    \centering
    \includegraphics[width=1\linewidth]{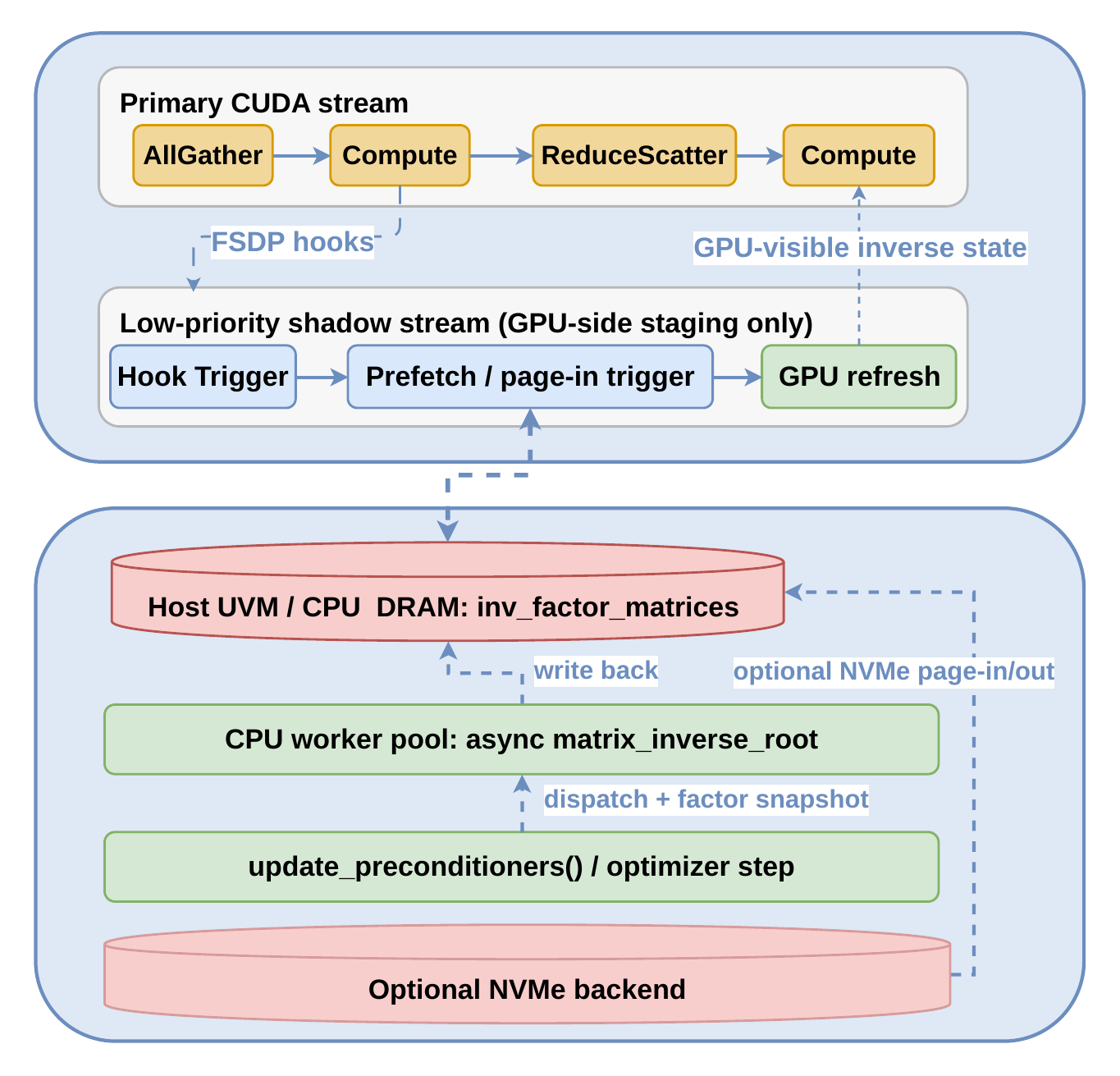}
    \caption{Shadow-stream staging and host-side asynchronous inverse-root updates in Asteria.}
    \label{fig:shadow-overlap-scheduling}
\end{figure}

\subsection{Hook-Driven Shadow Overlap Scheduling}
Moving large second-order states through an FSDP training pipeline increases pressure on shared interconnect and memory paths. Asteria addresses this problem with a runtime schedule that separates GPU-side state staging from host-side inverse-root computation.

Figure \ref{fig:shadow-overlap-scheduling} shows the resulting execution structure. The primary CUDA stream continues to execute the FSDP forward, backward, and communication phases. A low-priority shadow stream is used for GPU-side state staging, including prefetch and paging-related refresh. In parallel, a host-side worker pool executes asynchronous inverse-root updates on CPU snapshots of the factor matrices. After the host updates \texttt{inv\_factor\_matrices} in place, the runtime refreshes GPU-visible state so that subsequent preconditioning steps can consume the updated inverse factors.

\subsubsection{Hook-Based Scheduling Signals}
Asteria uses module-level PyTorch hooks as runtime scheduling signals. When UVM prefetch is enabled, it registers \texttt{register\_forward\_hook} to drain ready prefetch work. When backward JIT paging is enabled, it registers \texttt{register\_full\_backward\_pre\_hook} to schedule page-in for states that are likely to be needed soon in the backward pass. These hooks do not modify the main FSDP execution path. They provide points at which auxiliary state-staging work can be triggered.

\subsubsection{Shadow Stream and Host-Side Async Updates}
When a hook fires, Asteria schedules GPU-side staging work on a low-priority auxiliary CUDA stream rather than on the primary stream. This staging path is separate from the CPU worker path. The CPU worker pool computes inverse-root updates on host-side snapshots and writes the results back to host-resident \texttt{inv\_factor\_matrices}. After completed CPU jobs are collected, the runtime either queues the refreshed states for GPU-side prefetch or marks them dirty for NVMe paging management. Under UVM, Asteria refreshes CUDA-visible views and issues \texttt{cudaMemPrefetchAsync}. Under non-UVM settings, it refreshes GPU-side cached copies explicitly. When NVMe staging is enabled, the runtime can page state in before compute or use, and flush cold state back after collection. This design reduces exposed state-movement overhead while leaving the main FSDP training path unchanged.

\subsection{Topology-Aware Bounded-Staleness Coherence}
In multi-node training, Asteria does not treat second-order state synchronization as a flat global collective. Instead, it synchronizes host-resident preconditioner blocks through a topology-aware, bounded-staleness coherence path that matches the residency of \texttt{inv\_factor\_matrices}.

\subsubsection{Topology Cost Graph Discovery}
At initialization, Asteria builds a topology cost graph from distributed runtime information and optional NVML queries. The runtime first infers the local world size from environment variables such as \texttt{LOCAL\_WORLD\_SIZE}, with fallback to NVML or CUDA device count. It then groups ranks by node, assigns lower communication cost to intra-node edges and higher cost to inter-node edges, and refines intra-node links when NVML peer-to-peer information is available.

\subsubsection{Selective Coherence Registry}
Asteria maintains a \texttt{CoherenceRegistry} for second-order preconditioner blocks. For each block, the registry tracks a version counter and the step at which the block was last synchronized. During training, blocks whose coherence age remains within the configured staleness budget are treated as cache hits and skip communication. Only stale blocks enter the synchronization path.

\subsubsection{Hierarchical Host-Side Synchronization}
For blocks that exceed the staleness bound, Asteria synchronizes host-accessible tensors through topology-aware process groups. The runtime first averages a stale block within the node-local group, then averages it across node representatives, and finally broadcasts the refreshed block back to local peers. This communication path is implemented on host-side process groups so that CPU- or UVM-resident \texttt{inv\_factor\_matrices} can be synchronized directly without CPU-to-GPU-to-CPU round trips.

\subsubsection{Staleness Bound and Local Barrier}
Asteria also enforces a bounded-staleness barrier for asynchronous inverse-root computation. When the age of a local preconditioner version exceeds the configured bound while a pending CPU job is still in flight, the optimizer waits for completion before continuing. This barrier applies to local asynchronous compute progress and is separate from the topology-aware host-side coherence path.

\subsection{Implementation Details}
Asteria combines a Python runtime with a C++/CUDA extension exposed through \texttt{pybind11}. The extension implements UVM allocation, CUDA-visible host views, prefetch, page reclamation, and \texttt{io\_uring}-backed NVMe I/O, while the Python layer integrates these primitives with \texttt{AsyncDistributedShampoo}, asynchronous CPU workers, shadow staging, and topology-aware host-side coherence. The codebase also includes optional packed triangular SPD storage utilities.

\section{Experiments}

\subsection{Setup}
We evaluate Asteria in both scale-up and scale-out settings using the OLMo training framework~\cite{olmo20252olmo2furious}. The scale-up experiments target a memory-constrained single-node setting, while the scale-out experiments target distributed multi-node training under Fully Sharded Data Parallel (FSDP).

Across all experiments, we train OLMo-family language models from scratch on the English C4 corpus~\cite{raffel2023exploringlimitstransferlearning}, using a sequence length of 1024 tokens and the T5 tokenizer~\cite{raffel2023exploringlimitstransferlearning}. We study three model scales: a 660M model with \texttt{d\_model}=1408, 24 Transformer layers, and 22 attention heads; a OLMo-2 1B model with \texttt{d\_model}=2048, 24 layers, and 16 heads; and an OLMo-2 7B model with \texttt{d\_model}=4096, 32 layers, 32 attention heads, and \texttt{mlp\_hidden\_size}=22016. The 660M model follows the legacy OLMo configuration with GELU activations, whereas the 1B and 7B model follows the OLMo-2 configuration with SwiGLU and RMSNorm. All models use rotary positional embeddings (RoPE)~\cite{su2023roformerenhancedtransformerrotary} and omit bias terms.

We compare AdamW with second-order baselines including SOAP and KL-Shampoo under a shared FSDP training recipe. Unless otherwise noted, second-order runs use a precondition frequency of 10 and a maximum preconditioner dimension of 2048 \cite{shi2023distributeddataparallelpytorchimplementation,vyas2025soap}. For each optimizer and model scale, we select the learning rate from a sweep from $10^{-4}$ to $10^{-2}$  and report the run that achieves the lowest validation loss. Within each experiment group, we keep the sequence length, warmup schedule, and gradient clipping policy fixed, and we report optimizer-specific settings, including effective batch size and optimization hyperparameters, together with the corresponding results. In the memory-constrained scale-up setting, we enable the Asteria runtime features required for host-resident second-order state, staged state movement, and optional lower-tier storage.

\begin{figure}[htbp]
    \centering
    \includegraphics[width=1\linewidth]{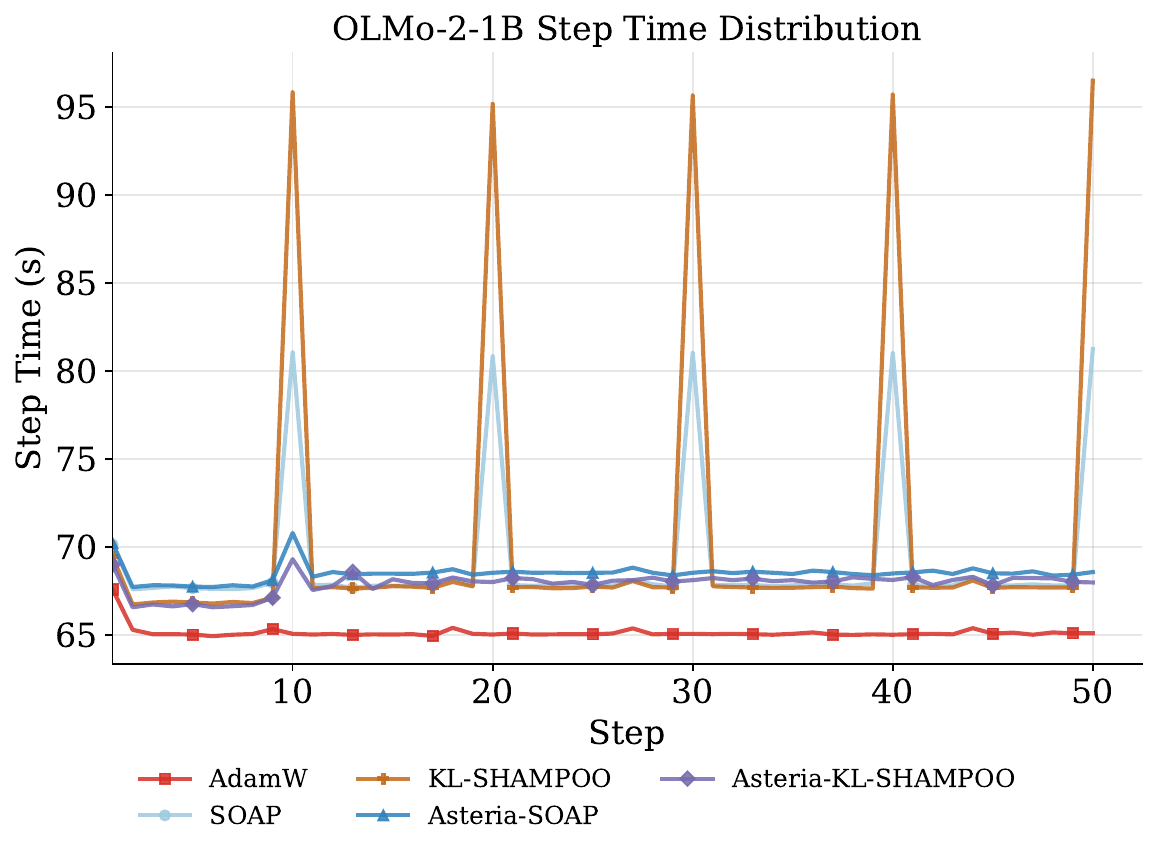}
    \caption{Step time distribution across training steps for OLMo-2-1B on the DGX Spark 
    }
    \label{fig:step_time_dist}
\end{figure}

\begin{figure}[htbp]
    \centering
    \includegraphics[width=1\linewidth]{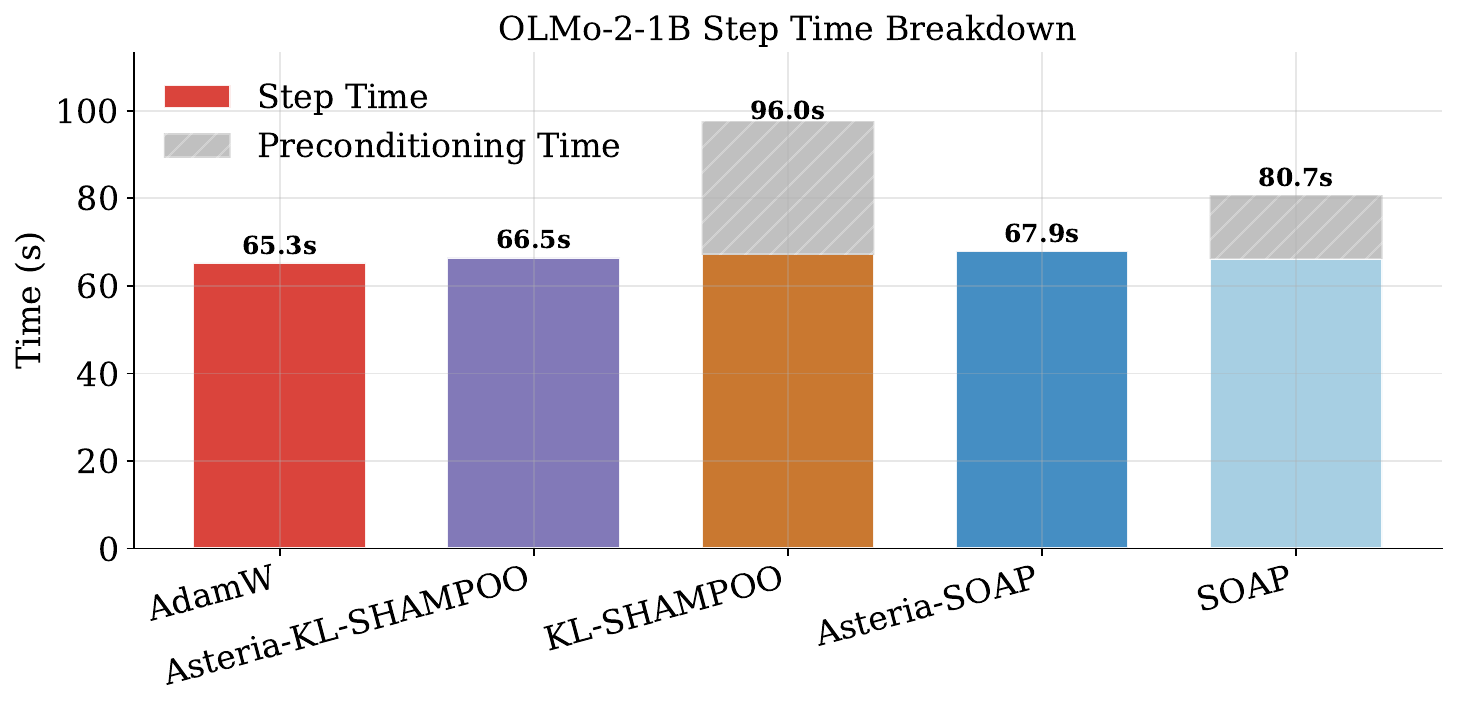}
    \caption{Step-time breakdown at the preconditioning boundary for OLMo-2-1B on the DGX Spark. 
    }
    \label{fig:step_time_breakdown}
\end{figure}

\begin{figure*}[h]
    \centering
    \includegraphics[width=1\linewidth]{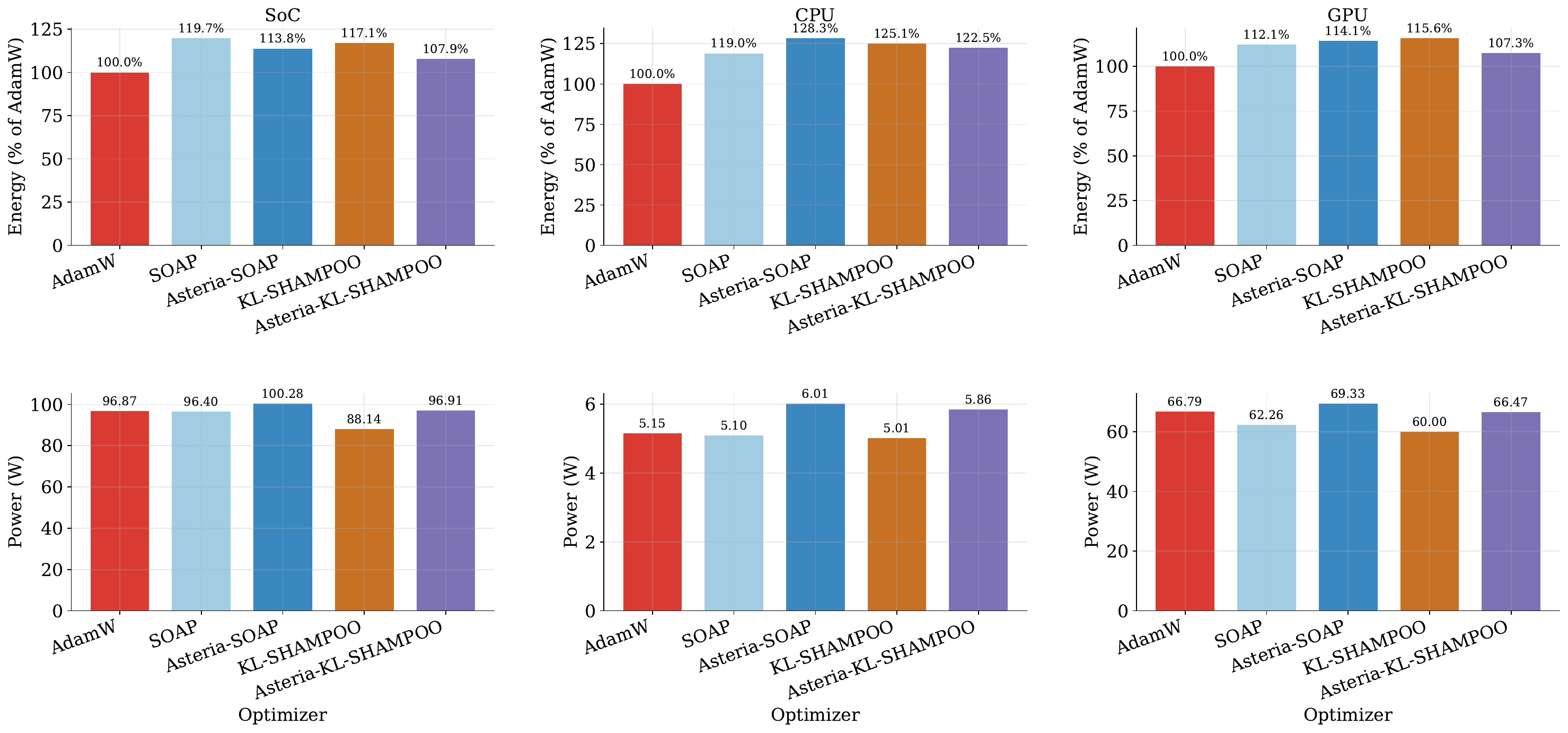}
    \caption{Energy and power comparison for OLMo-2-1B training for 50 steps on the DGX Spark. The top row reports total energy consumption for SoC, CPU, and GPU, normalized to the AdamW baseline (AdamW = 100\%). The bottom row reports the corresponding average power in watts.}
    \label{fig:energy-comparison-pf5-1b}
\end{figure*}

\subsection{Running on Nvidia DGX Spark}
\subsubsection{End-to-End Training Throughput and Latency Hiding}
To evaluate the system-level efficiency of Asteria, we profile the step-time dynamics during the pre-training of the OLMo-2-1B model. A fundamental drawback of native second-order optimizers is the severe periodic latency introduced by dense eigenvalue decompositions. As illustrated in Figure \ref{fig:step_time_dist}, native KL-SHAMPOO and SOAP experience massive step-time spikes at regular intervals (the precondition frequency is 10), peaking at 96.0s and 80.7s respectively. These synchronous blocking operations severely disrupt the training pipeline and degrade overall hardware utilization. 

In stark contrast, the integration of the Asteria runtime completely neutralizes these periodic spikes. Asteria-KL-SHAMPOO and Asteria-SOAP deliver a remarkably smooth and stable step-time distribution across the entire training trajectory. While Asteria introduces a negligible overhead of approximately 1.2 to 2.6 seconds per step compared to the purely first-order AdamW baseline (65.3s), it can preserve the near-constant throughput essential for later large-scale distributed training.

The mechanism behind this flat trajectory is explicitly captured in the step-time breakdown at the preconditioning boundary (Figure \ref{fig:step_time_breakdown}). In the native implementations, the preconditioning time (represented by the shaded grey regions) is fully exposed on the critical path, directly inflating the step time. Asteria, through its dual-DAG shadow scheduling and heterogeneous memory tiering, successfully decouples this $\mathcal{O}(N^3)$ computational burden. By pushing the inverse-root updates to background CPU workers and asynchronously percolating the updated states back to the GPU memory, Asteria effectively hides nearly 100\% of the preconditioning latency. Consequently, Asteria-KL-SHAMPOO completes the preconditioning step in just 66.5s, which is a dramatic reduction from the native 96.0s. This proves that Asteria can harvest the statistical efficiency of second-order optimization at a hardware cost nearly identical to first-order methods.In this setting, Asteria makes second-order optimization more energy-efficient than both its native counterparts and the AdamW baseline.

\subsubsection{Energy analysis}

\begin{figure}[h]
    \centering
    \includegraphics[width=0.75\linewidth]{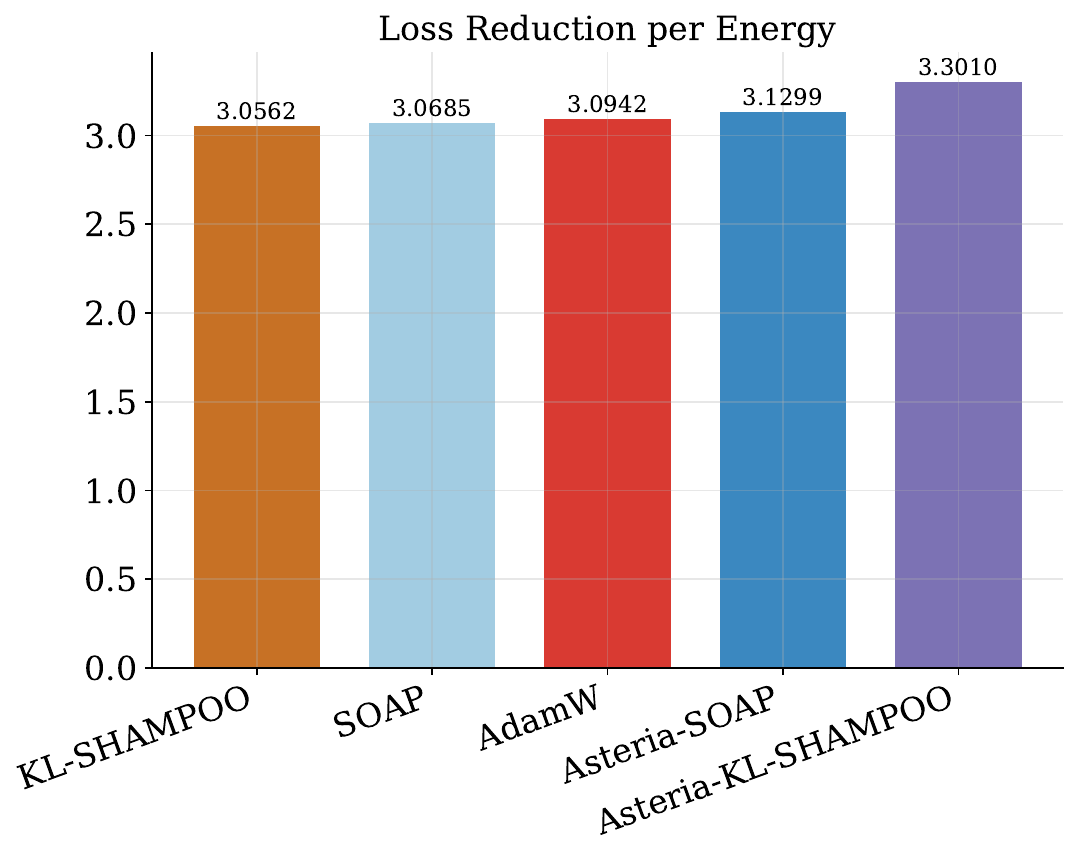}
    \caption{Energy-loss tradeoff for OLMo-2-1B training runs on the DGX Spark. Methods are ranked from lower to higher normalized loss-reduction efficiency, computed using Eq.~\eqref{eq:energy_loss_efficiency}. Higher values indicate better energy efficiency in converting energy into loss reduction.}
    \label{fig:energy-loss-tradeoff-pf5-1b}
\end{figure}

To accurately quantify the total and domain-specific energy consumption of workloads on the GB10 SoC, we utilized onboard hardware telemetry for both the CPU and GPU. By default, the GB10 platform does not expose the CPU power rails. To bypass this limitation, we interfaced with a discrete secondary MediaTek chipset on the motherboard. Using a custom Linux kernel module~\cite{spark_hwmon}, we queried an ACPI Device Specific Method (\_DSM) to expose the chip's hardware-level accumulating energy counters to the Linux hardware monitoring (hwmon) subsystem. While the standard Linux hwmon Application Binary Interface (ABI) dictates that energy counters report in increments of one microjoule (1.0 $\mu J$), empirical validation determined that the raw energy counters on this platform actually increment in units of 0.1 microjoules (0.1 $\mu J$). Energy consumption for each domain was calculated using the delta method: logging the respective counter states before and after workload execution, and dividing the difference by $10^7$ to yield values in standard Joules. This out-of-band approach ensures zero measurement overhead on the host CPU.

To validate the accuracy of this undocumented interface, we conducted a series of domain isolation tests, cross-referencing our measurements against standard NVIDIA Management Library (NVML) GPU telemetry~\cite{yang2024accurate} and the official 140W Thermal Design Power (TDP) limit of the GB10 SoC. Under an isolated CPU workload (stress-ng), the CPU energy counters scaled proportionally with utilization, while the GPU energy counters matched NVML's reported idle power draw. Conversely, under an isolated GPU workload (CUDA MatMul), the GPU energy counters matched NVML power draw levels, whereas the CPU energy counters accumulated at an empirically validated idle rate. Finally, under a combined maximal CPU and GPU stress test, the aggregate accumulation rate of both energy counters matched the 140W TDP ceiling of the GB10 SoC. This confirms that the telemetry accurately and independently captures both power domains.

For the energy-loss tradeoff analysis, we combined the measured total energy consumption with the final training loss of each run. Let $E_i$ denote the total energy consumed by method $i$, obtained by summing the per-domain energy deltas described above, and let $E_{\mathrm{AdamW}}$ denote the corresponding total energy of the AdamW baseline. Let $L_{\mathrm{final},i}$ denote the final loss of method $i$. Since all runs start from the same random initialization, we use a common reference loss $L_{\mathrm{init}}=\ln(V)\approx \ln(32128)\approx 10.377$, corresponding to the cross-entropy of a uniform next-token distribution under the T5 tokenizer vocabulary. We define the normalized loss-reduction efficiency of method $i$ as
\begin{equation}
\eta_i = \frac{L_{\mathrm{init}} - L_{\mathrm{final},i}}{E_i / E_{\mathrm{AdamW}}},
\label{eq:energy_loss_efficiency}
\end{equation}
where higher values indicate better energy efficiency, i.e., larger loss reduction achieved per unit normalized energy expenditure.

The system-level benefit of Asteria is reflected in both total energy and average power. As shown in the top row of Figure~\ref{fig:energy-comparison-pf5-1b}, native second-order optimizers incur a clear platform-level energy overhead over 50 training steps. Relative to AdamW, native SOAP consumes 119.7\% of the total SoC energy, while native KL-Shampoo consumes 117.1\%. Asteria reduces these totals to 113.8\% and 107.9\%, respectively, narrowing the energy gap to the first-order baseline.

A similar pattern appears when the energy is decomposed by device domain. On the CPU side, native SOAP consumes 119.0\% of AdamW's energy, whereas Asteria-SOAP increases CPU energy to 128.3\%. Native KL-Shampoo and Asteria-KL-Shampoo consume 125.1\% and 122.5\%, respectively. This increase in CPU energy for Asteria-SOAP is consistent with shifting more second-order work onto the host and keeping CPU workers active in the background. On the GPU side, however, Asteria produces a different trend. Native SOAP and KL-Shampoo consume 112.1\% and 115.6\% of AdamW's GPU energy, while Asteria-SOAP and Asteria-KL-Shampoo reduce these values to 114.1\% and 107.3\%, respectively. In particular, Asteria-KL-Shampoo lowers both GPU energy and total SoC energy relative to native KL-Shampoo, indicating that the reduction in exposed GPU idle time outweighs the additional host-side activity.

The bottom row helps explain this. Asteria does not reduce total energy by lowering instantaneous power draw. Instead, it shifts execution toward a higher-power but shorter-duration regime. For example, Asteria-KL-Shampoo increases average CPU power from 5.01\,W to 5.86\,W and average GPU power from 60.00\,W to 66.47\,W, while still reducing total SoC energy from 117.1\% to 107.9\%. These measurements are consistent with improved overlap between host-side inverse-root computation and GPU training, which improves end-to-end execution efficiency and hardware utilization.

Crucially, the system-level improvements of Asteria also translate into a better energy-loss tradeoff, as shown in Figure~\ref{fig:energy-loss-tradeoff-pf5-1b}. Native second-order optimizers do not outperform AdamW on this metric in the pf5 setting: KL-Shampoo and SOAP achieve normalized loss-reduction efficiencies of 3.0562 and 3.0685, both below AdamW at 3.0942. This indicates that their faster optimization progress per step is offset by the higher energy cost of second-order state maintenance. Asteria changes this tradeoff. Asteria-SOAP improves the metric to 3.1299, and Asteria-KL-Shampoo further increases it to 3.3010, which is the best result among all methods in this figure. These results show that reducing the exposed cost of second-order preconditioning improves not only runtime efficiency, but also the amount of loss reduction achieved per unit of energy. In this setting, Asteria turns second-order optimization from an energetically unfavorable choice into the most effective option in terms of loss reduction per energy. In this setting, Asteria makes second-order optimization more energy-efficient than both its native counterparts and the AdamW baseline.

\subsection{Running on Nvidia GH200 nodes}

\subsubsection{Convergence Efficiency and Staleness Tolerance}
\begin{figure}[htbp]
    \centering
    \includegraphics[width=1\linewidth]{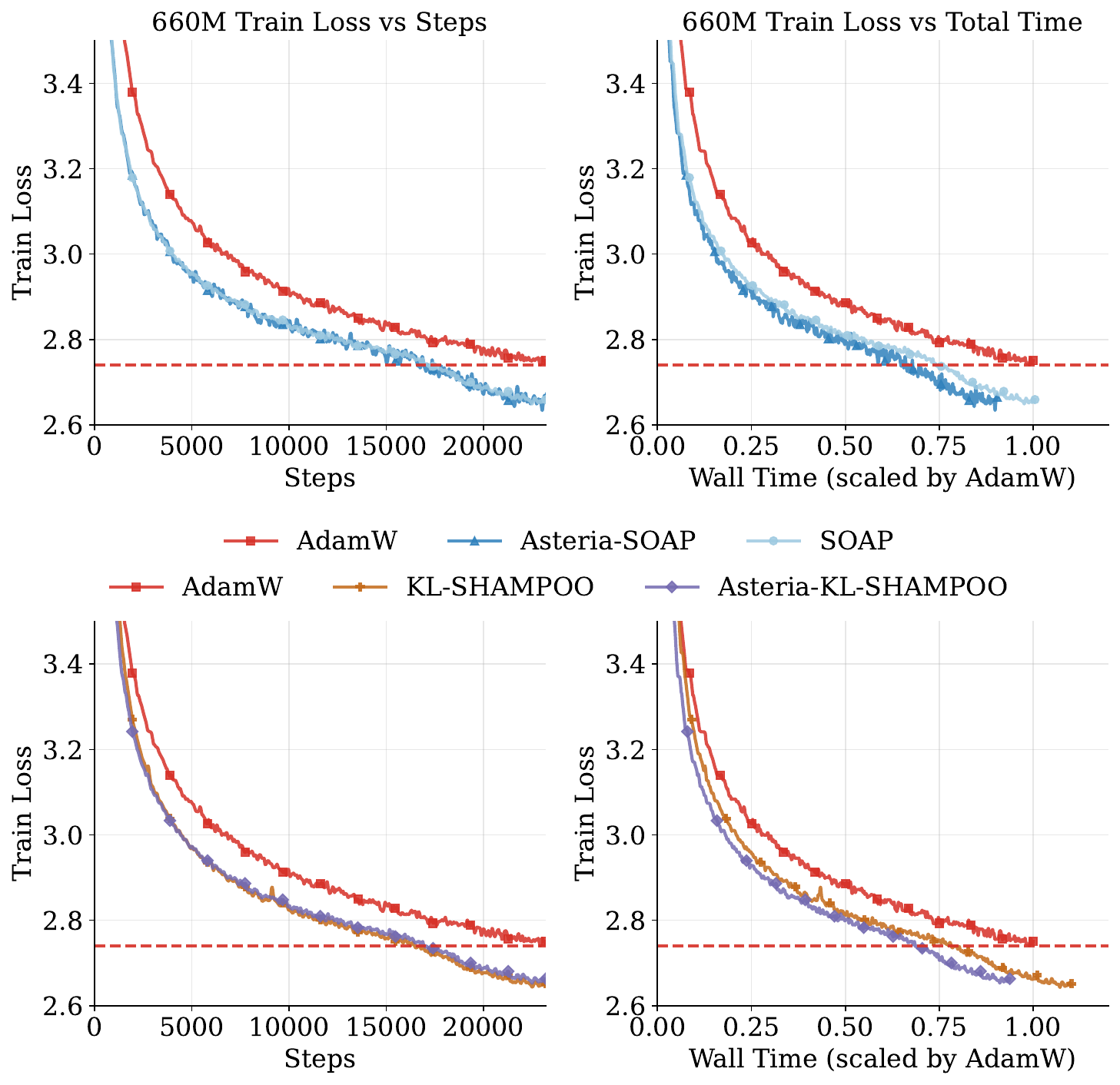}
    \caption{Training loss for 660M pretraining runs over optimizer steps (left column) and wall time normalized by AdamW (right column) on GH200 nodes.
    The dashed horizontal line marks the final training loss reached by AdamW. 
    Lower curves indicate faster convergence and better optimization efficiency.}
    \label{fig:660m-train-loss}
\end{figure}

To evaluate whether Asteria's asynchronous scheduling changes optimization behavior, we track the training loss of a 660M-parameter model throughout pretraining. Figure~\ref{fig:660m-train-loss} shows training loss over optimizer steps and normalized wall time.

The left column indicates that the Asteria variants remain close to their native second-order counterparts in step-wise convergence. Asteria-SOAP closely follows native SOAP, and Asteria-KL-SHAMPOO closely follows native KL-SHAMPOO over the full training trajectory. In both comparisons, the second-order methods reach lower training loss than AdamW, matching the findings in the previous literature \cite{lin2026understandingimprovingshampoosoap,vyas2025soap,eschenhagen2026clarifyingshampooadaptingspectral}. These results indicate that Asteria's bounded-staleness scheduling preserves the optimization behavior of the underlying second-order methods.

The right column shows the corresponding wall-time view. Relative to native SOAP and KL-SHAMPOO, the Asteria variants reach the same loss at lower normalized time. Both Asteria-SOAP and Asteria-KL-SHAMPOO cross the final AdamW loss level earlier than their native counterparts, as marked by the dashed horizontal line. This pattern is consistent with Asteria reducing the exposed cost of inverse-root computation by moving it off the critical path and overlapping host-side preconditioner updates with GPU training. As a result, Asteria retains the convergence benefit of second-order optimization while improving wall-clock efficiency.

\begin{figure}[htbp]
    \centering
    \includegraphics[width=1\linewidth]{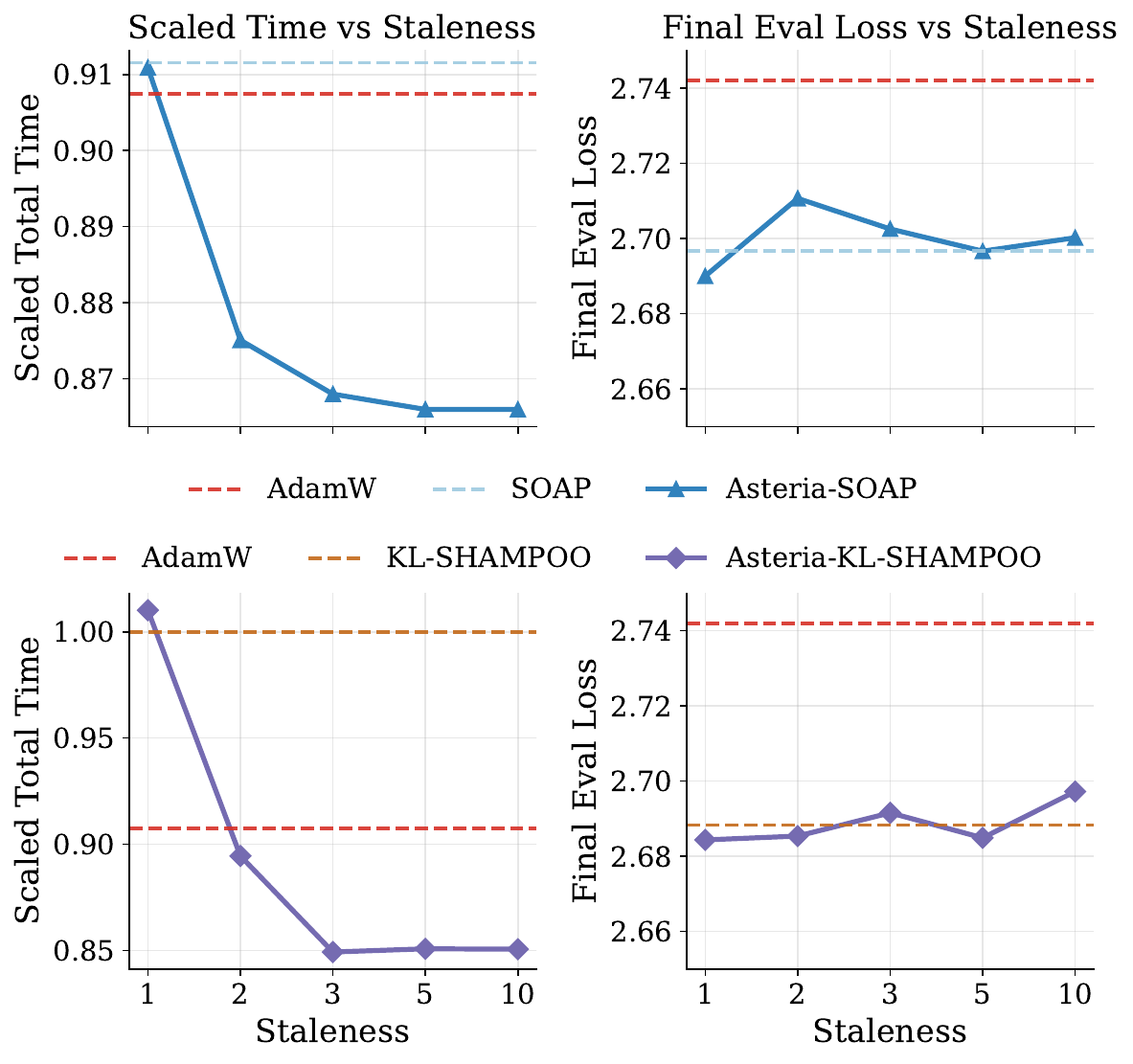}
\caption{Effect of asynchronous staleness on normalized training time (left) and final evaluation loss (right) for 660M pretraining runs on GH200 nodes. Dashed lines denote the AdamW and native second-order baselines. Lower values are better.}
    \label{fig:660m-staleness}
\end{figure}

To study the tradeoff between exposed system overhead and optimizer freshness, we vary the asynchronous staleness budget $S$ in Figure~\ref{fig:660m-staleness}. Here, $S$ denotes the maximum number of training steps for which the GPU may continue using an older preconditioner view while the CPU computes the updated inverse factors asynchronously.

The left column shows that small staleness leaves part of the second-order overhead exposed. At $S=1$, the total training time of Asteria-KL-SHAMPOO is slightly worse than that of native KL-SHAMPOO, and Asteria-SOAP remains close to the native SOAP baseline. This pattern suggests that a single GPU step does not provide enough execution overlap to hide the host-side inverse-root update. As $S$ increases to 3 and 5, total training time drops substantially for both optimizers and then plateaus, indicating that most of the exposed overhead has been hidden and that additional staleness provides little further benefit.

The right column shows that this runtime gain does not come with a corresponding loss in final model quality. For both Asteria-SOAP and Asteria-KL-SHAMPOO, the final evaluation loss varies only within a narrow range across $S \in \{1,2,3,5,10\}$ and remains close to the corresponding native second-order baseline. This result indicates that the bounded delay introduced by asynchronous preconditioner updates does not materially degrade optimization quality in this setting.

It is also important to distinguish system-level staleness $S$ from the algorithmic preconditioning frequency $pf$. Following established literature \cite{vyas2025soap,shi2023distributeddataparallelpytorchimplementation,lin2026understandingimprovingshampoosoap}, we keep $pf=10$ fixed across both native and Asteria runs and vary only $S$ . Based on the flat time curve beyond $S=3$ and the stable final evaluation loss, we use $S=5$ in the remaining experiments as a representative operating point that provides strong runtime benefit without observable degradation in final accuracy.

\subsubsection{Scaling Out: Multi-Node Efficiency on 1B and 7B Models}

\begin{figure}[htbp]
    \centering
    \includegraphics[width=1\linewidth]{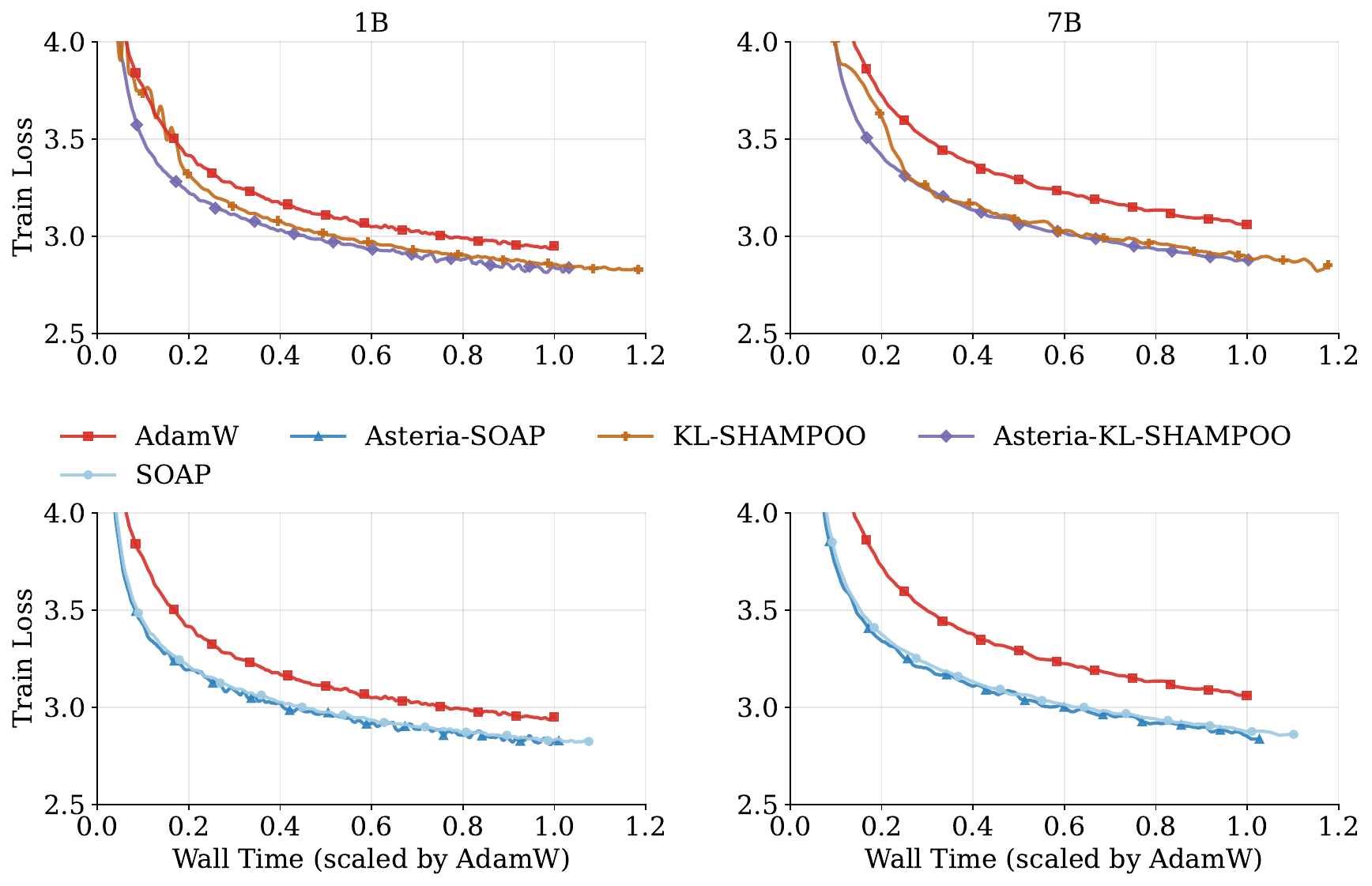}
    \caption{Training loss over wall time normalized by AdamW for 1B and 7B pretraining runs on GH200 nodes. The left column shows the 1B model and the right column shows the 7B model.  Lower curves indicate better time-to-convergence efficiency.}
    \label{fig:1b-7b-train-loss-time}
\end{figure}

To evaluate whether Asteria remains effective at larger model scales, we extend the comparison to 1B and 7B OLMo-2 models in a multi-node distributed setting. Because of the higher computational cost at these scales, the 1B and 7B pretraining runs are truncated at 10{,}000 and 5{,}000 steps, respectively.

Figure~\ref{fig:1b-7b-train-loss-time} shows training loss over normalized wall time for both model sizes. As model size increases, the cost of second-order state maintenance also increases, placing greater pressure on host-side inverse-root computation and distributed state movement. Despite this higher systems burden, the Asteria variants continue to preserve the wall-clock advantage of second-order optimization. For both the 1B and 7B models, Asteria-SOAP and Asteria-KL-SHAMPOO remain close to their native second-order counterparts while consistently outperforming AdamW in normalized time-to-convergence. This pattern is especially clear at 7B, where both Asteria variants maintain lower training loss than AdamW throughout the run and reach late-training loss levels earlier in normalized time. The close alignment between the Asteria and native second-order curves indicates that Asteria's runtime mechanisms do not materially degrade optimization behavior at a larger scale, while the persistent gap to AdamW shows that the convergence benefit of second-order methods is retained in multi-node training. Overall, these results show that Asteria preserves the convergence advantage of second-order methods while making that advantage usable in large-scale distributed training.

\subsubsection{Scaling Out: Strong-Scaling Efficiency}

\begin{figure}[htbp]
    \centering
    \includegraphics[width=1\linewidth]{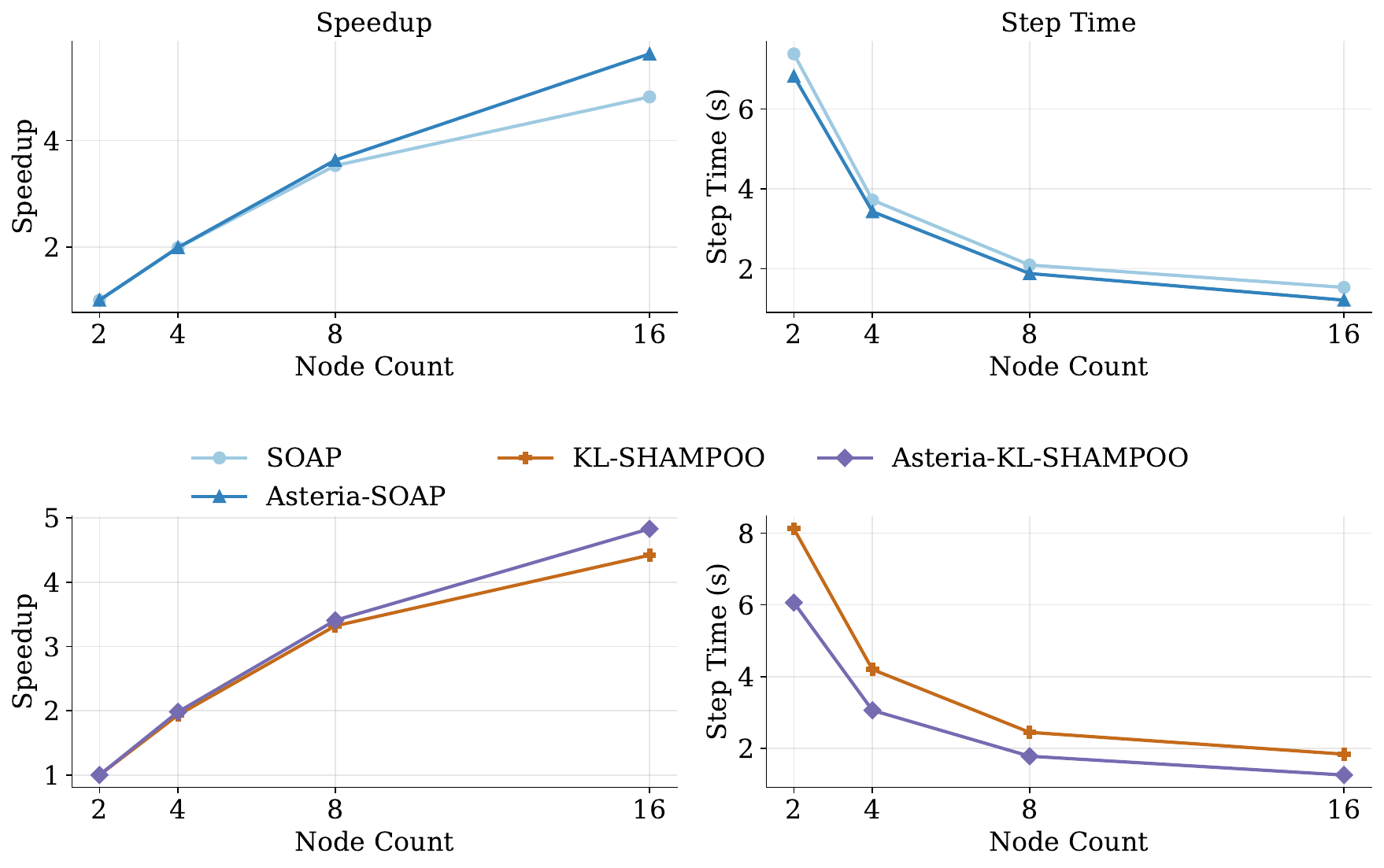}
    \caption{Strong-scaling results for 7B training with a fixed total workload on GH200 nodes. The left column shows speedup relative to the 2-node run, and the right column shows per-step execution time. Higher speedup and lower step time indicate better scaling efficiency.}
    \label{fig:strong-scaling-7b}
\end{figure}

To evaluate distributed scalability, we run strong-scaling experiments for the 7B model while increasing the node count from 2 to 16 at fixed total workload. This setting is challenging for second-order methods because per-device compute decreases with scale, while the relative cost of second-order state maintenance becomes larger.

Figure~\ref{fig:strong-scaling-7b} shows that Asteria improves strong-scaling behavior for both SOAP and KL-SHAMPOO. Across the full node range, Asteria achieves higher realized speedup and lower per-step time than the corresponding native baseline. The gain is especially clear for KL-SHAMPOO, where Asteria reduces step time substantially at every scale. These results indicate that Asteria reduces the exposed overhead of second-order state maintenance and preserves distributed efficiency more effectively as system scale increases.

\section{Discussion}

Our results suggest that the main obstacle to practical second-order optimization is not only the mathematical cost of curvature-aware updates, but also the systems mismatch between second-order state maintenance and modern training stacks. Native implementations can improve convergence in optimizer steps, yet their benefits are often obscured by exposed latency spikes, tight memory pressure, and synchronization overheads that are poorly aligned with how current accelerators and distributed runtimes operate. Asteria addresses this mismatch by decoupling expensive inverse-root updates from the latency-critical GPU path, placing different second-order states according to their asymmetric production and consumption patterns, and using bounded-staleness coordination to reduce unnecessary state movement. The resulting runtime preserves the optimization character of native second-order methods while making them substantially easier to sustain on real hardware.

The broader scientific significance is that these results recast second-order optimization as a co-design problem across algorithms, runtime scheduling, and heterogeneous memory systems. First, step-time profiling shows that much of the practical inefficiency of second-order methods comes from how their state maintenance is exposed to the execution pipeline, rather than from curvature-aware updates alone. Second, the single-node results show that large aggregated GPU memory is not a fundamental prerequisite for second-order training: with lifecycle-aware placement and staged access, second-order state can be supported within much tighter device-visible memory budgets. Third, the energy measurements indicate that better orchestration improves not only throughput, but also physical efficiency, allowing more loss reduction per unit energy. Finally, the multi-node experiments suggest that distributed second-order training benefits from treating preconditioner freshness and communication as runtime-managed resources rather than rigidly synchronized global state.

Taken together, these findings point to a broader conclusion: whether second-order methods are practical at LLM scale depends as much on systems design as on optimizer mathematics. By showing that state placement, asynchronous compute, and selective synchronization can be jointly engineered, Asteria helps narrow the long-standing gap between the statistical promise of second-order optimization and its real-world deployability.

Our evaluation focuses on unified-memory systems and GH200-based distributed clusters, which are precisely the settings where second-order optimization is most exposed to memory pressure and distributed state-management overhead. The fact that Asteria remains effective in these regimes highlights the value of treating optimizer-state placement, asynchronous compute, and synchronization as a joint systems problem. While more constrained discrete-GPU platforms with narrow PCIe links and additional parallel training regimes such as tensor or pipeline parallelism remain to be explored, the current results establish that the core runtime principles of Asteria already provide substantial benefits in two demanding environments: memory-limited single-node training and bandwidth-sensitive multi-node scaling.

\section{Conclusion}

We presented Asteria, a runtime for practical second-order optimization under memory and communication constraints. By combining asymmetric memory tiering, asynchronous host-side inverse-root computation, hook-assisted GPU-side staging, and bounded-staleness selective coherence, Asteria keeps second-order state maintenance largely off the critical path. Across single-node and multi-node experiments, Asteria makes large second-order runs feasible under tight memory budgets, preserves the convergence behavior of native second-order methods, improves time-to-convergence, and yields better energy-loss tradeoffs. These results show that second-order optimization can become a practical systems choice for LLM training when optimizer-state placement, asynchronous compute, and synchronization are co-designed with the underlying hardware.

\bibliographystyle{IEEEtran}
\bibliography{references}

\begin{thebibliography}{10}
\providecommand{\url}[1]{#1}
\csname url@samestyle\endcsname
\providecommand{\newblock}{\relax}
\providecommand{\bibinfo}[2]{#2}
\providecommand{\BIBentrySTDinterwordspacing}{\spaceskip=0pt\relax}
\providecommand{\BIBentryALTinterwordstretchfactor}{4}
\providecommand{\BIBentryALTinterwordspacing}{\spaceskip=\fontdimen2\font plus
\BIBentryALTinterwordstretchfactor\fontdimen3\font minus \fontdimen4\font\relax}
\providecommand{\BIBforeignlanguage}[2]{{%
\expandafter\ifx\csname l@#1\endcsname\relax
\typeout{** WARNING: IEEEtran.bst: No hyphenation pattern has been}%
\typeout{** loaded for the language `#1'. Using the pattern for}%
\typeout{** the default language instead.}%
\else
\language=\csname l@#1\endcsname
\fi
#2}}
\providecommand{\BIBdecl}{\relax}
\BIBdecl

\bibitem{loshchilov2019decoupledweightdecayregularization}
\BIBentryALTinterwordspacing
I.~Loshchilov and F.~Hutter, ``Decoupled weight decay regularization,'' 2019. [Online]. Available: \url{https://doi.org/10.48550/arXiv.1711.05101}
\BIBentrySTDinterwordspacing

\bibitem{shampoo_raw}
\BIBentryALTinterwordspacing
V.~Gupta, T.~Koren, and Y.~Singer, ``Shampoo: Preconditioned stochastic tensor optimization,'' in \emph{Proceedings of the 35th International Conference on Machine Learning}, ser. Proceedings of Machine Learning Research, J.~Dy and A.~Krause, Eds., vol.~80.\hskip 1em plus 0.5em minus 0.4em\relax PMLR, 10--15 Jul 2018, pp. 1842--1850. [Online]. Available: \url{https://doi.org/10.48550/arXiv.1802.09568}
\BIBentrySTDinterwordspacing

\bibitem{vyas2025soap}
\BIBentryALTinterwordspacing
N.~Vyas, D.~Morwani, R.~Zhao, I.~Shapira, D.~Brandfonbrener, L.~Janson, and S.~M. Kakade, ``{SOAP}: Improving and stabilizing shampoo using adam for language modeling,'' in \emph{The Thirteenth International Conference on Learning Representations}, 2025. [Online]. Available: \url{https://doi.org/10.48550/arXiv.2409.11321}
\BIBentrySTDinterwordspacing

\bibitem{eschenhagen2026clarifyingshampooadaptingspectral}
\BIBentryALTinterwordspacing
R.~Eschenhagen, A.~Cai, T.-H. Lee, and H.-J.~M. Shi, ``Clarifying shampoo: Adapting spectral descent to stochasticity and the parameter trajectory,'' 2026. [Online]. Available: \url{https://doi.org/10.48550/arXiv.2602.09314}
\BIBentrySTDinterwordspacing

\bibitem{lin2026understandingimprovingshampoosoap}
\BIBentryALTinterwordspacing
W.~Lin, S.~C. Lowe, F.~Dangel, R.~Eschenhagen, Z.~Xu, and R.~B. Grosse, ``Understanding and improving shampoo and soap via kullback-leibler minimization,'' 2026. [Online]. Available: \url{https://doi.org/10.48550/arXiv.2509.03378}
\BIBentrySTDinterwordspacing

\bibitem{shi2023distributeddataparallelpytorchimplementation}
\BIBentryALTinterwordspacing
H.-J.~M. Shi, T.-H. Lee, S.~Iwasaki, J.~Gallego-Posada, Z.~Li, K.~Rangadurai, D.~Mudigere, and M.~Rabbat, ``A distributed data-parallel pytorch implementation of the distributed shampoo optimizer for training neural networks at-scale,'' 2023. [Online]. Available: \url{https://doi.org/10.48550/arXiv.2309.06497}
\BIBentrySTDinterwordspacing

\bibitem{osawa2023asdlunifiedinterfacegradient}
\BIBentryALTinterwordspacing
K.~Osawa, S.~Ishikawa, R.~Yokota, S.~Li, and T.~Hoefler, ``Asdl: A unified interface for gradient preconditioning in pytorch,'' 2023. [Online]. Available: \url{https://doi.org/10.48550/arXiv.2305.04684}
\BIBentrySTDinterwordspacing

\bibitem{zhao2023pytorchfsdpexperiencesscaling}
\BIBentryALTinterwordspacing
Y.~Zhao, A.~Gu, R.~Varma, L.~Luo, C.-C. Huang, M.~Xu, L.~Wright, H.~Shojanazeri, M.~Ott, S.~Shleifer, A.~Desmaison, C.~Balioglu, P.~Damania, B.~Nguyen, G.~Chauhan, Y.~Hao, A.~Mathews, and S.~Li, ``Pytorch fsdp: Experiences on scaling fully sharded data parallel,'' 2023. [Online]. Available: \url{https://doi.org/10.48550/arXiv.2304.11277}
\BIBentrySTDinterwordspacing

\bibitem{martens2020optimizingneuralnetworkskroneckerfactored}
\BIBentryALTinterwordspacing
J.~Martens and R.~Grosse, ``Optimizing neural networks with kronecker-factored approximate curvature,'' 2020. [Online]. Available: \url{https://doi.org/10.48550/arXiv.1503.05671}
\BIBentrySTDinterwordspacing

\bibitem{anil2021scalablesecondorderoptimization}
\BIBentryALTinterwordspacing
R.~Anil, V.~Gupta, T.~Koren, K.~Regan, and Y.~Singer, ``Scalable second order optimization for deep learning,'' 2021. [Online]. Available: \url{https://doi.org/10.48550/arXiv.2002.09018}
\BIBentrySTDinterwordspacing

\bibitem{lu2026meanfisherorthogonalprojectionnatural}
\BIBentryALTinterwordspacing
Y.~Lu and W.~Armour, ``Beyond the mean: Fisher-orthogonal projection for natural gradient descent in large batch training,'' 2026. [Online]. Available: \url{https://doi.org/10.48550/arXiv.2508.13898}
\BIBentrySTDinterwordspacing

\bibitem{ren2021zerooffloaddemocratizingbillionscalemodel}
\BIBentryALTinterwordspacing
J.~Ren, S.~Rajbhandari, R.~Y. Aminabadi, O.~Ruwase, S.~Yang, M.~Zhang, D.~Li, and Y.~He, ``Zero-offload: Democratizing billion-scale model training,'' 2021. [Online]. Available: \url{https://doi.org/10.48550/arXiv.2101.06840}
\BIBentrySTDinterwordspacing

\bibitem{rajbhandari2021zeroinfinitybreakinggpumemory}
\BIBentryALTinterwordspacing
S.~Rajbhandari, O.~Ruwase, J.~Rasley, S.~Smith, and Y.~He, ``Zero-infinity: Breaking the gpu memory wall for extreme scale deep learning,'' 2021. [Online]. Available: \url{https://doi.org/10.48550/arXiv.2104.07857}
\BIBentrySTDinterwordspacing

\bibitem{Fang_2023}
\BIBentryALTinterwordspacing
J.~Fang, Z.~Zhu, S.~Li, H.~Su, Y.~Yu, J.~Zhou, and Y.~You, ``Parallel training of pre-trained models via chunk-based dynamic memory management,'' \emph{IEEE Transactions on Parallel and Distributed Systems}, vol.~34, no.~1, p. 304–315, Jan. 2023. [Online]. Available: \url{https://doi.org/10.48550/arXiv.2108.05818}
\BIBentrySTDinterwordspacing

\bibitem{kingma2017adammethodstochasticoptimization}
\BIBentryALTinterwordspacing
D.~P. Kingma and J.~Ba, ``Adam: A method for stochastic optimization,'' 2017. [Online]. Available: \url{https://doi.org/10.48550/arXiv.1412.6980}
\BIBentrySTDinterwordspacing

\bibitem{shazeer2018adafactoradaptivelearningrates}
\BIBentryALTinterwordspacing
N.~Shazeer and M.~Stern, ``Adafactor: Adaptive learning rates with sublinear memory cost,'' 2018. [Online]. Available: \url{https://doi.org/10.48550/arXiv.1804.04235}
\BIBentrySTDinterwordspacing

\bibitem{shoeybi2020megatronlmtrainingmultibillionparameter}
\BIBentryALTinterwordspacing
M.~Shoeybi, M.~Patwary, R.~Puri, P.~LeGresley, J.~Casper, and B.~Catanzaro, ``Megatron-lm: Training multi-billion parameter language models using model parallelism,'' 2020. [Online]. Available: \url{https://doi.org/10.48550/arXiv.1909.08053}
\BIBentrySTDinterwordspacing

\bibitem{10.5555/3488766.3488792}
\BIBentryALTinterwordspacing
Y.~Jiang, Y.~Zhu, C.~Lan, B.~Yi, Y.~Cui, and C.~Guo, ``A unified architecture for accelerating distributed {DNN} training in heterogeneous {GPU/CPU} clusters,'' in \emph{14th USENIX Symposium on Operating Systems Design and Implementation (OSDI 20)}.\hskip 1em plus 0.5em minus 0.4em\relax USENIX Association, Nov. 2020, pp. 463--479. [Online]. Available: \url{https://dl.acm.org/doi/10.5555/3488766.3488792}
\BIBentrySTDinterwordspacing

\bibitem{Maurya_2024}
\BIBentryALTinterwordspacing
A.~Maurya, J.~Ye, M.~M. Rafique, F.~Cappello, and B.~Nicolae, ``Deep optimizer states: Towards scalable training of transformer models using interleaved offloading,'' in \emph{Proceedings of the 25th International Middleware Conference}, ser. Middleware ’24.\hskip 1em plus 0.5em minus 0.4em\relax ACM, Dec. 2024, p. 404–416. [Online]. Available: \url{https://doi.org/10.48550/arXiv.2410.21316}
\BIBentrySTDinterwordspacing

\bibitem{stich2019localsgdconvergesfast}
\BIBentryALTinterwordspacing
S.~U. Stich, ``Local sgd converges fast and communicates little,'' 2019. [Online]. Available: \url{https://doi.org/10.48550/arXiv.1805.09767}
\BIBentrySTDinterwordspacing

\bibitem{daily2018gossipgradscalabledeeplearning}
\BIBentryALTinterwordspacing
J.~Daily, A.~Vishnu, C.~Siegel, T.~Warfel, and V.~Amatya, ``Gossipgrad: Scalable deep learning using gossip communication based asynchronous gradient descent,'' 2018. [Online]. Available: \url{https://doi.org/10.48550/arXiv.1803.05880}
\BIBentrySTDinterwordspacing

\bibitem{olmo20252olmo2furious}
\BIBentryALTinterwordspacing
T.~OLMo, P.~Walsh, L.~Soldaini, D.~Groeneveld, K.~Lo, S.~Arora, A.~Bhagia, Y.~Gu, S.~Huang, M.~Jordan, N.~Lambert, D.~Schwenk, O.~Tafjord, T.~Anderson, D.~Atkinson, F.~Brahman, C.~Clark, P.~Dasigi, N.~Dziri, A.~Ettinger, M.~Guerquin, D.~Heineman, H.~Ivison, P.~W. Koh, J.~Liu, S.~Malik, W.~Merrill, L.~J.~V. Miranda, J.~Morrison, T.~Murray, C.~Nam, J.~Poznanski, V.~Pyatkin, A.~Rangapur, M.~Schmitz, S.~Skjonsberg, D.~Wadden, C.~Wilhelm, M.~Wilson, L.~Zettlemoyer, A.~Farhadi, N.~A. Smith, and H.~Hajishirzi, ``2 olmo 2 furious,'' 2025. [Online]. Available: \url{https://doi.org/10.48550/arXiv.2501.00656}
\BIBentrySTDinterwordspacing

\bibitem{raffel2023exploringlimitstransferlearning}
\BIBentryALTinterwordspacing
C.~Raffel, N.~Shazeer, A.~Roberts, K.~Lee, S.~Narang, M.~Matena, Y.~Zhou, W.~Li, and P.~J. Liu, ``Exploring the limits of transfer learning with a unified text-to-text transformer,'' 2023. [Online]. Available: \url{https://doi.org/10.48550/arXiv.1910.10683}
\BIBentrySTDinterwordspacing

\bibitem{su2023roformerenhancedtransformerrotary}
\BIBentryALTinterwordspacing
J.~Su, Y.~Lu, S.~Pan, A.~Murtadha, B.~Wen, and Y.~Liu, ``Roformer: Enhanced transformer with rotary position embedding,'' 2023. [Online]. Available: \url{https://doi.org/10.48550/arXiv.2104.09864}
\BIBentrySTDinterwordspacing

\bibitem{spark_hwmon}
\BIBentryALTinterwordspacing
A.~Kapenekakis, ``spark\_hwmon,'' 2026, accessed 2026-04-08. [Online]. Available: \url{https://github.com/antheas/spark\_hwmon}
\BIBentrySTDinterwordspacing

\bibitem{yang2024accurate}
\BIBentryALTinterwordspacing
Z.~Yang, K.~Adamek, and W.~Armour, ``Accurate and convenient energy measurements for gpus: A detailed study of nvidia gpu’s built-in power sensor,'' in \emph{SC24: International Conference for High Performance Computing, Networking, Storage and Analysis}.\hskip 1em plus 0.5em minus 0.4em\relax IEEE, 2024, pp. 1--17. [Online]. Available: \url{https://doi.org/10.1109/SC41406.2024.00028}
\BIBentrySTDinterwordspacing

\end{thebibliography}

\end{document}